\documentclass[fleqn,usenatbib]{mnras}

\usepackage{newtxtext,newtxmath}

\usepackage[T1]{fontenc}
\usepackage{ae,aecompl}
\usepackage{pifont}
\newcommand{\cmark}{\text{\ding{51}}}
\newcommand{\xmark}{\text{\ding{55}}}

\usepackage{graphicx}	
\usepackage{amsmath}	
\usepackage[dvipsnames]{xcolor}
\usepackage[normalem]{ulem}
\usepackage{comment}

\title[Realistic sensitivity maps for DM detection]{Detecting low-mass haloes with strong gravitational lensing I: the effect of data quality and lensing configuration}
\author[Giulia Despali et al.]{Giulia Despali $^{1}$\thanks{E-mail:gdespali@uni-heidelberg.de}, Simona Vegetti$^{2}$, Simon D. M. White$^{2}$, Devon M. Powell$^{2}$, 
Hannah R.  Stacey$^{2}$, \newauthor
Christopher D. Fassnacht$^{3}$, Francesca Rizzo$^{4}$, Wolfgang Enzi$^{2}$\\
\\
$^{1}$ Zentrum für Astronomie der Universität Heidelberg, Institut für Theoretische Astrophysik, Albert-Ueberle-Str. 2, 69120 Heidelberg\\
$^{2}$ Max Planck Institute for Astrophysics, Karl-Schwarzschild-Strasse 1, 85748 Garching bei M{\"u}nchen, Germany\\
$^{3}$ Department of Physics and Astronomy, University of California, Davis, 1 Shields Ave., Davis, CA 95616, USA\\
$^{4}$Cosmic Dawn centre (DAWN), University of Copenhagen, Lyngbyvej 2, DK-2100 Copenhagen, Denmark\\
}

\date{Accepted 2021 December 1. Received 2021 November 28; in original form 2021 July 30}

\pubyear{2021}

\begin{document}
\label{firstpage}
\pagerange{\pageref{firstpage}--\pageref{lastpage}}
\maketitle

\begin{abstract}
This paper aims to quantify how the lowest halo mass that can be detected with galaxy-galaxy strong gravitational lensing depends on the quality of the observations and the characteristics of the observed lens systems. Using simulated data, we measure the lowest detectable NFW mass at each location of the lens plane, in the form of detailed \emph{sensitivity maps}. In summary, we find that: (i) the lowest detectable mass $M_{\rm low}$ decreases linearly as the signal-to-noise ratio (SNR) increases and the sensitive area is larger when we decrease the noise; (ii) a moderate increase in angular resolution (0.07" vs 0.09") and pixel scale (0.01" vs 0.04") improves the sensitivity by on average 0.25 dex in halo mass, with more significant improvement around the most sensitive regions; (iii) the sensitivity to low-mass objects is largest for bright and complex lensed galaxies located inside the caustic curves and lensed into larger Einstein rings (i.e $r_{E}\geq1.0"$). We find that for the sensitive mock images considered in this work, the minimum mass that we can detect at the redshift of the lens lies between $1.5\times10^{8}$ and $3\times10^{9}M_{\odot}$. We derive analytic relations between $M_{\rm low}$, the SNR and resolution and discuss the impact of the lensing configuration and source structure. Our results start to fill the gap between approximate predictions and real data and demonstrate the challenging nature of calculating precise forecasts for gravitational imaging. In light of our findings, we discuss possible strategies for designing strong lensing surveys and the prospects for HST, Keck, ALMA, Euclid and other future observations. 

\end{abstract}

\begin{keywords}
(cosmology:) dark matter -- Cosmology -- gravitational lensing: strong -- galaxies: high-redshift -- cosmology: observations -- methods: data analysis
\end{keywords}



\section{Introduction}

Strong gravitational lensing is one of the most promising methods for studying the nature of dark matter. It allows one to detect low-mass dark haloes within the haloes of lens galaxies and along their line of sight, providing a quantitative test of the Cold Dark Matter (CDM) paradigm in a halo mass regime that is not accessible to any other technique. It offers a robust method to distinguish between CDM and alternative models in which the abundance of low-mass haloes is suppressed, for example, Warm Dark Matter \citep[WDM, e.g.][]{schneider12,lovell12,lovell14}, Fuzzy Dark Matter \citep[FDM, e.g.][]{robles17} and Self-Interacting Dark Matter \citep[SIDM, e.g.][]{vogel12,vogel16,despali19}. 

In the past few years, the search for dark matter haloes and subhaloes through their effect on magnified arcs and Einstein rings gained much attention and led to the detection of a few low-mass dark matter clumps using optical Hubble Space Telescope \citep[HST,][]{vegetti10,vegetti10b} observations, Keck adaptive optics imaging \citep{vegetti12} or interferometric data from the Atacama Large Millimeter/submillimeter Array \citep[ALMA,][]{hezaveh16}. The distortion is due to gravity only, allowing one to directly measure the mass of the object acting as a lens independently of its baryonic content, both in the case of the main lens galaxy (typically an Early-Type galaxy) and additional smaller perturbers. 

A significant improvement towards constraining dark matter with lensing has been achieved by considering the contribution of low-mass haloes located along the line of sight (i.e. \emph{field haloes}), in addition to that of substructures in the lens. For configurations in which both the source and lens are at a large cosmological distance, \citet{li16b} and \citet{despali18} demonstrated that these isolated low-mass haloes represent the dominant contribution to the number of detectable objects \citep[see also][]{metcalf05,gilman18}. 
Recently, \citet{amorisco21} and \citet{he21} investigated the lensing by \emph{field} haloes further, including the degeneracies with the main lens model and a scatter in the concentration-mass relation of haloes, finding that these can have additional (and opposite) effects on the total number of detectable objects.

Despite these theoretical improvements, observational results coming from optical data have not set yet strong constraints on the nature of dark matter. Homogeneous samples of lens systems have been used to measure the (sub)halo mass function, combining the information coming both from detections and non-detections \citep{vegetti14,ritondale19b} and found it to be consistent with expectations from the CDM paradigm in the regime that these samples could probe. The current best optical samples of galaxy-galaxy lens systems only allow one to rule out WDM models with thermal relic mass $m_{\rm DM}$<2keV \citep{vegetti18,ritondale19b,birrer17,ritondale19a}. Another lensing technique able to constrain dark matter is the analysis of flux-ratio anomalies in distant lensed quasars \citep{metcalf01,xud15,nierenberg14,gilman19,hsueh19}: current limits are more stringent than those derived from gravitational imaging and exclude WDM models with $m_{\rm DM}<5.2$ keV. Independent constraints at a similar level come from other probes such as the Lyman-$\alpha$ forest \citep[e.g.][]{irsic17,murgia18} and the Milky Way satellite count \citep[][]{jethwa16,newton21} - see \citet{enzi21} for a comprehensive list of recent studies using all these techniques. Joint analyses are able to exclude WDM models with particle masses $m_{\rm DM}<6.048$ keV \citep{enzi21} or $m_{\rm DM}<9.7$ keV \citep{nadler21}. However, future higher-resolution observations or larger samples will improve the sensitivity of lensed arcs and provide constraints on  dark matter at a level comparable to the other probes.

In practice, obtaining reliable and precise forecasts for 
the number and properties of gravitational lens systems that are required to obtain more stringent constraints is very challenging. Intuitively, it is evident that observational data of increasing quality - in terms of signal-to-noise ratio and angular resolution - will allow us to obtain stronger constraints. 
For example, increasing the angular resolution allows one to see perturbations on smaller scales and thus to detect lower mass haloes. However, several other factors influence the results in addition to the angular resolution of the instrument: the redshift of the lens and source galaxies, the details of the surface brightness distribution of the source, the size of the lensed images, the noise level of the data and the detection threshold used to define a detection. In this work, we attempt to fill the gap between approximate predictions and real data by quantifying these effects. We hope that the results of this paper will help to design a strategy for future strong lensing surveys targeted at constraining dark matter through the detection of low-mass haloes. To this end, we generate mock lensing data with properties mimicking those of observed systems, simulating the lensed images at different signal-to-noise ratios (SNR) and angular resolution. We then analyse these mock data-sets as we would real data and calculate the data sensitivity, i.e. the range of halo masses that the data allow us to detect in CDM. We discuss the relative advantages and disadvantages of $(i)$ increasing the size of the observed samples, $(ii)$ increasing the signal-to-noise level while keeping the resolution fixed (i.e. through longer exposure times), and $(iii)$ re-observing a small number of systems with higher resolution instruments. We consider observations with the Hubble Space Telescope, Keck Adaptive Optics, ALMA and Euclid and discuss prospects for other future observations.

In this first paper we discuss the detectability of perturbers located at the redshift of the main lens (i.e. subhaloes), focusing on estimates of the detectable masses and the dependence on the lensing configuration, data quality and source structure. In the second paper of this series, we will expand our results by including field haloes located along the line of sight and calculate the total number of detectable perturbers. In that work, we will also discuss how many lenses we would need to distinguish CDM from WDM, while here we focus on predictions from CDM only.

The paper is structured as follows: in Section \ref{method} we present the lens modelling and the method used to estimate the sensitivity, together with the adopted theoretical framework for the distribution and properties of the perturbers. In Section \ref{mocks} we then describe the mock data created and analysed for this work. In Section \ref{results} we present the analysis of the sensitivity maps and quantify the effect of the signal-to-noise ratio and the angular resolution of the data, while in Section \ref{configuration} we discuss how the properties of the source and the lensing configuration can affect the results.
These results are then used in Section \ref{other_data} to discuss the data quality and the properties of the observed systems that would maximise the chances of constraining dark matter with future lensing observations. Finally, we summarise our results and draw our conclusions in Section \ref{concl}. We discuss additional systematic errors and other sources of bias that could affect the forecast and the lensing analysis in general in the Appendix.

\section{Methodology} 
\label{method}

This section describes our model for the structure of the lens galaxies and the abundance, structure and lensing effects of the perturbing low-mass haloes. We then describe how we calculate observational sensitivity maps for each considered system, by which we mean a quantitative characterisation of the potential for detecting a low-mass halo at each point in the field through the perturbations it induces in the images of the background source. Our results will demonstrate that an accurate sensitivity map is a {\it sine qua non} for inferring constraints on structure formation models from the observations.

\subsection{Lens modelling}

In the context of Bayesian statistics, the strong lensing inference problem is best expressed in terms of the following posterior distribution:
\begin{equation}
    P(\mathbf{s},\boldsymbol{\eta},\lambda_s,|\mathbf{d})=\frac{P(\mathbf{d}|\mathbf{\eta},\mathbf{s})P(\boldsymbol{\eta})P(\mathbf{s}|\lambda_s)}{P(\mathbf{d})}.
\label{eq_post}    
\end{equation}
Here, $\mathbf{d}$ is the observed surface brightness distribution of the lensed images, $\mathbf{s}$ the background-source-galaxy surface brightness, $\boldsymbol{\eta}$ a vector containing the parameters describing the lens mass distribution, and $\lambda_{s}$ the source regularisation level. We represent the lens mass distribution by an elliptical power-law model, with dimensionless surface mass density given by
\begin{equation}
    \kappa(x,y)=\frac{\kappa_{0}\left(2-\frac{\gamma}{2}\right)q^{\gamma-3/2}}{2\left(q^{2}(x^{2}+r_{c}^{2})+y^{2}\right)^{(\gamma-1)/2}}\,.
    \label{eq_lens}
\end{equation}
Hence, the unknown parameters $\boldsymbol{\eta}$ include the normalisation $\kappa_{0}$, the radial mass-density slope $\gamma$, the axis ratio $q$ (and position angle). In addition, an external shear component of strength $\Gamma$ and position angle $\Gamma_{\theta}$ is added to the model. The core radius $r_c$ is fixed at $10^{-4}$ arcsec. For an isothermal spherical lens $\kappa_{0}$ can be interpreted as the Einstein radius. We follow  \citet{vegetti09} and model the source $\mathbf{s}$ in a pixellated regularised fashion. The brightness of each pixel in the source plane, as well as the source regularisation level $\lambda_{s}$, are thus also free parameters of the model. We refer the reader to the original paper and \citet{rybak15}, \citet{rizzo18} and \citet{powell21} for more details on the lens modelling procedure. 

In this paper, we mainly focus on the Bayesian evidence and its relation to the sensitivity of the data to the presence of low-mass haloes. In particular, we are interested in comparing the evidence of a smooth model (i.e. a lens without perturbing subhaloes)
\begin{equation}
 E_{\rm smooth} = \int P(\mathbf{d}|\boldsymbol{\eta},\mathbf{s}) ~P(\mathbf{s}|\lambda_{s})~P(\boldsymbol{\eta})~d\lambda_{s}dsd\boldsymbol{\eta}\,,
 \label{eq_ev_smooth}
\end{equation}
with that of a model including a subhalo 
\begin{equation}
E_{\rm pert} = \int P(\mathbf{d}|\boldsymbol{\eta}, \boldsymbol{\eta}_{\rm{sub}})~P(\mathbf{s}|\lambda_{s})~P(\boldsymbol{\eta})~d\lambda_{s} ds d\mathbf{\eta}\,.
 \label{eq_ev_pert}
\end{equation}
Here $\boldsymbol{\eta_{\rm sub}}$ contains the parameters describing a subhalo located at a certain projected position $\mathbf{x}$; its structural properties can be expressed as a function of mass and concentration as it is normally done for \emph{field} haloes - $\boldsymbol{\eta}_{\rm{sub}} = \{m,c,\mathbf{x}\}$ - or, alternatively, as a function of the subhalo maximum circular velocity $V_{\rm max}$ and the radius $r_{\rm max}$ at which this velocity is attained \citep{springel08b} - $\boldsymbol{\eta}_{\rm{sub}} = \{V_{\rm max},r_{\rm max},\boldsymbol{x}\}$.

Both integrals are performed with {\rm MultiNest}  \citep{feroz13} assuming uniform priors on the parameters $\boldsymbol{\eta}$ and a uniform prior in logarithmic space on the source regularisation $\lambda_{s}$. We choose the size of the priors to be the same for the two models. Below, we discuss how we use the Bayes factor $\Delta\log E = \log  E_{\rm smooth} -  \log E_{\rm pert}$ to quantify the lowest detectable halo mass. 

\subsection{Sensitivity map}
\label{method_sens}

For our purposes,  sensitivity maps define the lowest detectable (sub)halo mass $M_{\rm low}(\mathbf{x},c)$ at each location on the lens plane. These maps are the principal analysis tool for interpreting the detection (or non-detection) of low-mass haloes with strong lensing observations: they allow us to calculate the expected number of detectable (sub)haloes in a pixel, in a specific lens system, or in a sample of lens systems. For each observed system, the sensitivity depends on: (i) the source and lens redshift and their relative positions on the sky; (ii) the mass density profile of the main lens; (iii) the amount of structure in the surface brightness distribution of the source at the observed wavelength; (iv) the effective point-spread function and the signal-to-noise ratio achieved by the observational set-up; (v) the threshold chosen to accept a detection; and (vi) also (weakly) on cosmology.

We start by defining the lowest detectable subhalo mass $M_{\rm low}(\mathbf{x},c)$ at a given position $\mathbf{x}$ on the lens plane as the mass of an NFW subhalo that, when added to the main lens, creates a  difference in the Bayes factor between the smooth and perturbed models of $\Delta\log E \geq50$ (see equations \ref{eq_ev_smooth} and \ref{eq_ev_pert}). Having created the data with a smooth lens (see Section \ref{mocks}), $M_{\rm low}$ is, at each position, the lowest subhalo mass whose presence is ruled out by the data at a robust statistical level. 
In practice, we create a two-dimensional grid in subhalo mass and position and for each point on this grid we perform the integral (\ref{eq_ev_pert}) to find the lowest mass satisfying our detection criteria. Given the quality of the data used here, we consider masses between $10^{8}$ and $10^{11}M_{\odot}$. 

We describe the perturbing subhaloes with NFW profiles \citep{navarro96},
\begin{equation}
\rho(r)=\dfrac{\rho_{s}r_s}{r\left(1+r/r_s\right)^{2}}\,,
\label{eq_lens_nfw}
\end{equation}
where $\rho(r)$ is the spherically averaged density as a function of radius,  $r_s$ is the scale-radius, and $\rho_s$ is the density normalisation. The typical values of $\rho_s$ and $r_{s}$ for haloes of given mass at each redshift are here set by the (redshift-dependent) concentration-mass relation \citep{duffy08}. This means that here we fix the subhalo concentration to the mean value $\bar{c}(m)$ predicted by the relation for any given mass at the redshift of the lens $z=z_{L}$. As a result we have: $\boldsymbol{\eta_{\rm sub}}= \{m,c=\bar{c}(m),\mathbf{x}\}$ and $M_{\rm low}=M_{\rm low}(\mathbf{x},c=\bar{c}(m))$.

The (lensing) properties of subhaloes normally do not follow the same scaling relations as isolated haloes; however, this is not crucial at this stage, since the aim of this paper is to characterise the relative variations in sensitivity from one system to the other. Moreover, this choice of profile will allow us, in the second paper of this series, to compare the properties of perturbers at $z=z_{L}$ to those located along the line of sight \citep{despali18} more easily.

Recent works \citep{minor21a,minor21b} point to the fact that not all dark sources of perturbations can be well described by NFW profiles, and this could be especially relevant for subhaloes. We plan to explore this in future work.

Our approach is similar to those used by \citet{vegetti14} and \citet{ritondale19b} but with an important difference. Unlike them, we marginalise both over the source (including its regularisation level) and over the main lens parameters.  As discussed in more detail in Appendix \ref{app_mm}, we have found this marginalisation to be important to account fully for the degeneracy between the main lens and perturber properties, especially when the latter are described by NFW profiles. Failing to allow readjustment of the main lens potential can lead to overly optimistic predictions for the sensitivity: a large difference in Bayesian evidence arising from a failure to reproduce the image positions can be compensated by a small change in the ({\it a priori} unknown) main lens parameters without requiring a superposed small halo. Recently, \citet{amorisco21} and \citet{he21} also found that varying the main lens potential is essential, and that doing so is even more important when considering NFW \emph{field} haloes as perturbers rather than subhaloes. We will address this issue in the second paper of this series.

The threshold $\Delta\log E \geq50$ is chosen because it roughly corresponds to a 10-$\sigma$ detection threshold \citep{ritondale19b} and has been established as a reliable way to limit the rate of false-positive detections, and so to provide robust and conservative results \citep{ritondale19b}. In Appendix \ref{sens_lower} we also discuss how the sensitivity map changes when a lower detection threshold, similar to the one adopted by \citet{hezaveh16}, is used instead.

\begin{table*}
    \begin{center}

    \begin{tabular}{ccccccccccc}
    \hline
    System & Acronym &$z_{\rm L}$&$z_{\rm S}$ & $R_{E}$ & Instrument & pixel size & PSF FWHM  & Camera & Band & $\lambda$\\
     &  & & & [arcsec] &  & [arcsec] & [arcsec]  &  &  & \\
    \hline
    J1110+3649  & BELLS1 & 0.733 & 2.502 & 1.141 & HST & 0.04 &0.09 & WFC3 & F606W & 588.7 nm\\
    J1201+4743 & BELLS2 & 0.563 & 2.126 & 1.035 & HST & 0.04 & 0.09 & WFC3 &F606W & 588.7 nm\\
    JVAS\,B1938+666 & SHARP1 &0.881 & 2.059 & 0.4156 & Keck-AO & 0.01 &0.07 & NIRC2 &K'& 2200 nm\\
    SPT\,0532$-$50 & ALMA1 & 1.15 & 3.399 & 0.557 & ALMA & - & 0.045 & -& band 7 & 880 $\mu$m\\
    \hline
    \end{tabular}
    \end{center}
    \caption{Summary of real lens systems used to create mock lensing observations: ID of the system, acronym assigned in the course of this paper for simplicity, lens and source redshift, and Einstein radius. We then list the original instrument configuration also used in the creation of the mocks, the pixel size and  angular resolution (defined by the FHWM of the PSF) in units of arc seconds, the band and wavelength of the observation \citep{ritondale19a,vegetti12}. The PSF models are shown in Figure \ref{psf_image}. The HST data were taken with one orbit of observing time, corresponding to $\sim$45 min, as quoted in Table \ref{tab_mocks}.}
    \label{tab_obs}
\end{table*}
\begin{table*}
    \begin{center}

    \begin{tabular}{ccccccccc}
      & & & &Observation setup & & \\
    \hline
     name & source & lens & FHWM  & pixel size & observing  & SNR & $R_{E}$  & realistic\\
      &  model & model  &  [arcsec] &  [arcsec] & time & (median) & [arcsec] & configuration \\
    \hline
    M1 & BELLS1 & BELLS1 & 0.09 & 0.04 & 45 mins & 3.5 & 1.141  & \cmark\\
    M2 & BELLS1 & BELLS1 & 0.09 & 0.04& 3 hours & 6.13 & 1.141  & \cmark\\
    M3 & BELLS1 & BELLS1 & 0.09 & 0.04& 6.5 hours & 9.24 & 1.141  & \cmark\\
    M4 & BELLS2 & BELLS2 & 0.09 & 0.04& 45 min& 4.5 &1.035  & \cmark\\
    M5 & BELLS2 & BELLS2 & 0.09 & 0.04& 3 hours & 8.55 & 1.035 & \cmark\\
    M6 & BELLS2 & BELLS2 & 0.09 & 0.04& 6.5 hours & 12.89 &1.035 & \cmark\\
    \hline
    M7 & double Gaussian & BELLS1  & 0.09 & 0.04& 45 mins & 2.1 &  1.141 & \xmark\\
    M8 & bright Gaussian & BELLS1 & 0.09 & 0.04 & 45 mins & 5.8 &  1.141  &\xmark\\
    M9 & NGC 5457 & BELLS1 & 0.09& 0.04 & 45 mins & 5.6 &  1.141 & \xmark\\
    \hline
    M10 & SHARP1 & SHARP1 & 0.09 & 0.04 & - & 4.1*& 0.4156 &\xmark\\
    M11 & SHARP1 - shifted & SHARP1 & 0.09 & 0.04 & - & 4.7* &0.4156  &\xmark\\
    M12 & NGC 5457 & SHARP1 & 0.09& 0.04 & - & 11.7*& 0.4156  &\xmark\\
    M13 & NGC 5457 & BELLS1 & 0.09 & 0.04 & - & 21*&  1.141 &\xmark\\
    \hline
    M14 & SHARP1 & SHARP1 & 0.07 & 0.01 & 4 hours& 4.1& 0.4156  & \cmark\\
    M15 & SHARP1 - shifted & SHARP1 & 0.07 & 0.01 & 4 hours & 4.7& 0.4156 &  \cmark\\
    M16 & NGC 5457 & SHARP1 & 0.07 & 0.01 & - & 11.7 & 0.4156  &\xmark\\
    M17 & NGC 5457 & BELLS1 & 0.07& 0.01 & - & 21&  1.141  &\xmark\\
    \hline
    M18 & BELLS1 & BELLS1 & 0.16 & 0.1&$\sim$45 mins  & 2&  1.141 &  \cmark\\
    M19 & NGC 5457 & BELLS1 & 0.16 & 0.1& - & 12 & 1.141  &\xmark\\
    \hline
    M20 & ALMA1 & ALMA1 & 0.045 & - & 4 hours & 4 & 0.557 &\cmark\\
    \hline
    \end{tabular}
    \end{center}
    \caption{Summary of mock data sets used in this work. We list the combination of source and lens model (from table \ref{tab_obs}), PSF model and pixel size, angular resolution, observing time and the SNR, calculated as the median value per pixel - except for the mock images M10-M13, where the SNR is matched to that of M14-M17 per unit area (here we quote the same SNR value for the two cases, marking the matched one with *). Finally, we list the size of the Einstein radius $R_{E}$ and in the last column, we mark the mock images which fully resemble realistic observations in terms of the combination of all the properties listed so far.}
    \label{tab_mocks}
\end{table*}

\section{Mock data} \label{mocks}

The goal of this paper is a systematic study of how the observational setup influences the sensitivity function: to this end, we vary the angular resolution and SNR of the observations, the source structure and the lensing configuration. Doing so with actual data would require many new observations, while this work aims instead to help the design of future observational strategies or the interpretation of existing observational results. We chose, therefore, to work with mock data so that we can investigate the effect of each variable in a controlled fashion while keeping uncertainties and systematic errors under control. We discuss potential limitations to this approach throughout the paper.
In this section, we describe the properties of our sample of simulated data. Table \ref{tab_obs} summarises the adopted lens models and the corresponding original observational set-up; the mock images are summarised instead in Table \ref{tab_mocks}.

\subsection{Source and lens galaxies}

Even though we are working with simulated data, we still want to ensure that we are considering realistic observational parameters. To this end, we select two gravitational lens systems from the BELLS-GALLERY sample \citep[][- GO: 14189; PI: Bolton]{shu16a} and one from the SHARP survey \citep[][- Keck Program ID: 2010A-U085N2L; PI: Fassnacht]{lagattuta12}.  For simplicity, we refer to these systems as BELLS1, BELLS2 and SHARP1 throughout the paper - see Table \ref{tab_obs}. 
The  BELLS-GALLERY sample consists of HST (WCF3, F606W) observations of high-redshift ($\bar{z}\sim3$) Lyman-alpha emitters lensed by ($\bar{z}\sim1$) massive Early-Type galaxies. The lens modelling of this sample by \citet{ritondale19a}  revealed compact and clumpy sources surrounded by elongated filamentary structures. 
The SHARP survey focuses on Keck-II AO (NIRC2,  K' band) observations of red lensed galaxies. Here, we consider the gravitational lens system JVAS B1938+666 \citep{vegetti12}, a bright infrared galaxy at redshift 2.059 gravitationally lensed into an almost complete Einstein ring, by a massive galaxy at $z\sim 0.8$. Initially discovered at radio frequencies \citep{king97}, the NIR observations reveal a rather round and smooth  
source galaxy \citep{lagattuta12,vegetti12}.

\citet{ritondale19a} and \citet{vegetti12}  have modelled these systems assuming a pixellated model for the background source galaxy and an elliptical power-law mass density profile for the foreground lens galaxy (see equation \ref{eq_lens}). We start from their most probable reconstructed sources and lens them forward through their best smooth lens model (i.e. without the contribution of subhaloes), creating mock observations with the same set-up of the original data.
In order to explore the effect of source structure further, we also use three additional source models, combined with the BELLS1 lens model: two analytical sources created with Gaussian distributions and the image of a low-$z$ spiral galaxy (NGC 5457) extracted from HST archival data (Proposal ID:13361, PI: Blair). For the latter, we rescale the size appropriately to the chosen redshift of the source and the range of surface brightness to match that of the best reconstructed sources from the real data-sets. 

We create simulated observations by convolving the lensed images with a realistic PSF and by adding instrumental noise. In total, we create 21 strong gravitational lens systems of varying configuration, SNR and angular resolution (see Table \ref{tab_mocks}). Mock images M1-M6 also contain a S\'ersic model of the lens light distribution from the HST analysis, while we have neglected it in the other cases. We find that the inclusion of the lens light, when correctly modelled, does not alter our results.

\subsection{PSF model} 
\label{sec:psf_models} 
 
\begin{figure}
     \centering
    \includegraphics[width=0.9\hsize]{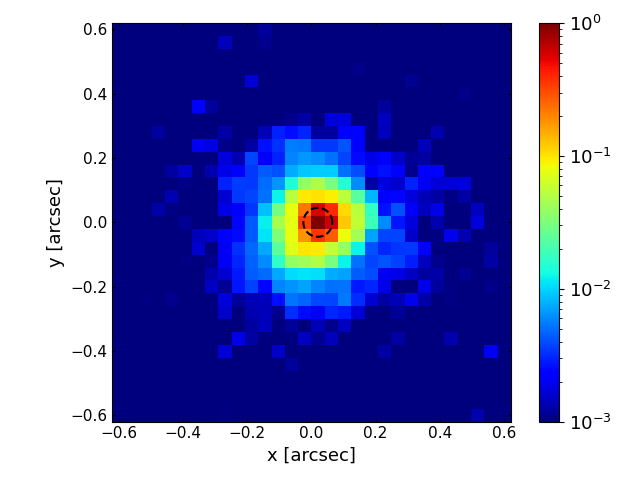}
    \includegraphics[width=0.9\hsize]{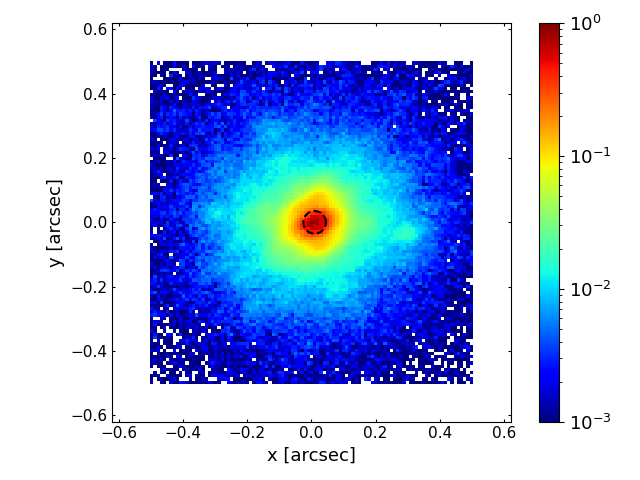}
    
\caption{PSF models used in this work and taken from the analysis of the original obsserved systems: the BELLS-GALLERY lenses \citep{ritondale19a} observed with HST-WFC3 (top) and the system B1938+666 \citep{vegetti12} observed with Keck-AO NIRC2 (bottom). The black dashed circle shows the region enclosed by one FHWM, which determines the angular resolution: 0.09 arcsec in the top panel and 0.07 arcsec in the bottom panel. The corresponding wavelengths and bands are listed in Table \ref{tab_obs}. }
    \label{psf_image}
\end{figure}

The exact radial shape of the PSF depends on the wavelength of the observations and, in the case of adaptive optics, also on the specific observing conditions as well as the efficiency of the AO system \citep{ragland18}. Thus, to create images as realistic as possible, we adopt the PSF models used in the analysis of the original HST-WCF3 F606W and Keck-II NIRC2 K-band data (see Figure \ref{psf_image}). The corresponding angular resolution, given by the FHWM of the PSF, is 0.09 and 0.07 arcsec, while the pixel scale is 0.04 and 0.01 arcsec, respectively. The angular resolution and pixel scale of the mock images are listed in Table \ref{tab_mocks}: the mock data sets M$_1$ to M$_{13}$ have an angular resolution of 0.09 arcsec and a pixel scale of 0.04 arcsec; for the lens systems, M$_{14}$ to M$_{17}$ we adopt a resolution of 0.07 arcsec and a pixel scale of 0.01 arcsec. Finally, the simulated data M$_{18}$ and M$_{19}$ are characterised by a PSF FWHM of 0.16 arcsec and a pixel scale of 0.1 arcsec (expected to be representative of Euclid VIS observations).

\subsection{Noise model} 
\label{sec:noise_models} 

For each pixel, we consider the contribution of a Poisson component and a Gaussian one, with a standard deviation proportional to $\sqrt{t_{\rm obs}}$. As a consequence, the overall SNR scales as $\sqrt{t_{\rm obs}}$.
Again, we use existing observations to make an informed decision on our choice of the observing time and the resulting noise level. For the three lens models selected for this work, the median SNR of the original data (measured as median SNR per pixel) is between 3.5 and 4.5, making them comparable in this respect. We modify the noise in our simulated data according to the needs of our experiment:
\begin{enumerate}

\item[--] in the two sets of data M1-M2-M3 and M4-M5-M6, the SNR is progressively increased from this original values (M1 and M4), to simulate an increase in observing time and investigate its effect on the data sensitivity;
\item[--] for the models M7-M9, we use the original Gaussian noise value from M1;
\item[--] M14 and M15 share the original noise value from M13;
\item[--] we create the two groups of mocks M14-M17 and M10-M13 so that, in each corresponding pair, the SNR {\it per unit area} (rather than per pixel) is the same, so that the effects of angular resolution (both pixel size and FWHM) can be investigated separately from those due to photon statistics. Given the different pixel size in each pair of data set, 16 pixels in the higher resolution image cover the same area the area in arcsec$^{2}$ of one pixel in the lower one - we thus use this ratio to match the SNR across the pairs of images. In practice, this means that the SNR per pixel is lower in the higher resolution case of each pair. One limitation of this approach is that, in reality, an increase in resolution requires a change of instrument, or filter, leading to source galaxies with a different structure. 

\end{enumerate}

\begin{figure*}
     \centering
    \includegraphics[width=\hsize]{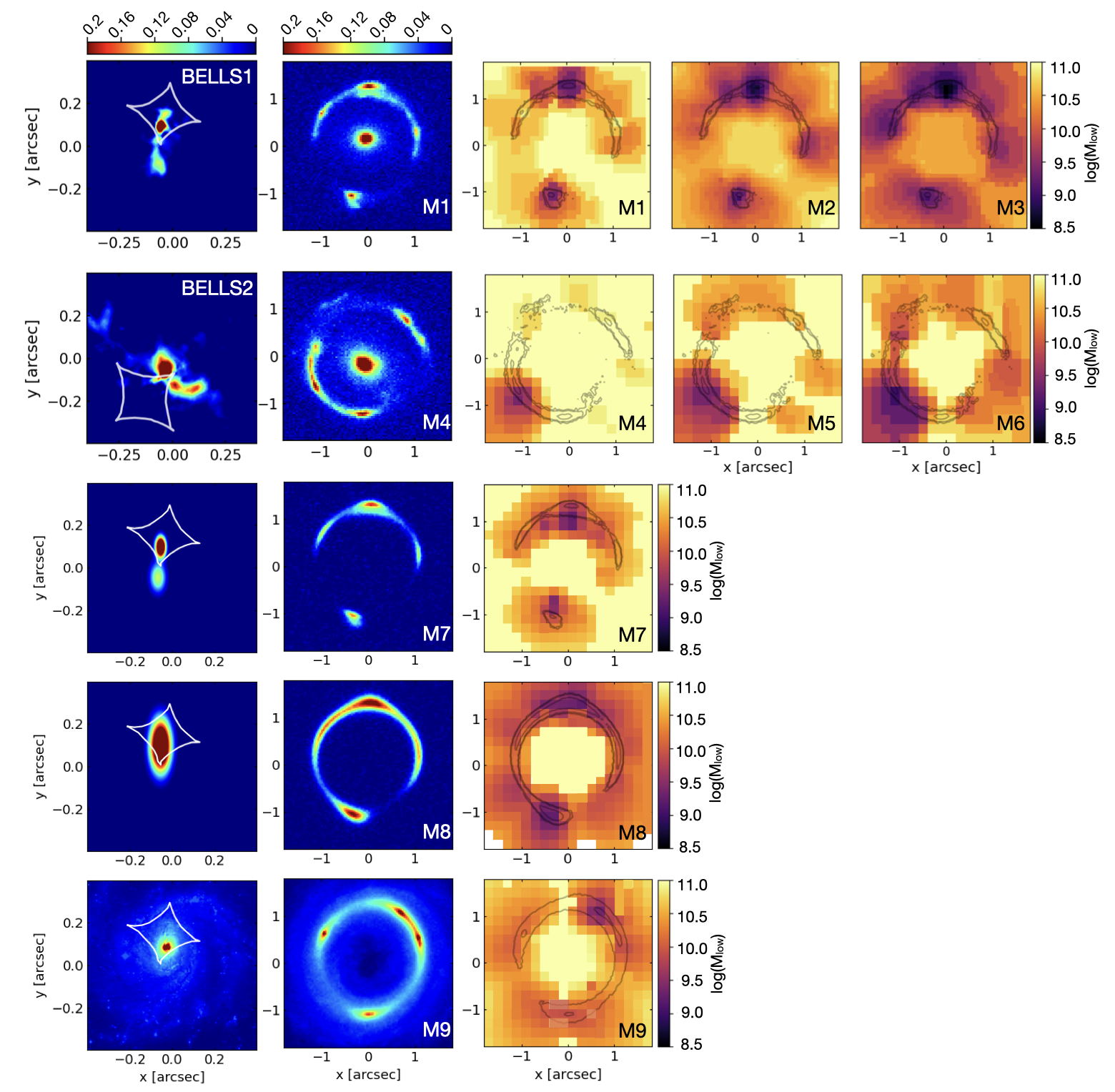}
    
\caption{Summary plot showing the mock data sets M1 to M9, simulated at a resolution of 0.09 arcsec. From left to right we show the source model, the lensed images and the corresponding sensitivity maps for each case. The caustics lines are shown in white on the source plane. In the top two rows, the images are based on two BELLS-GALLERY systems  from \citet{ritondale19a} and reproduce the original lensing configuration at the original or improved SNR; these also contain the lensed images together with a S\'ersic model of the galaxy light. Here we only show the mock images at the original level of SNR (M1 and M4), while the sensitivity maps are also shown for the mock images where the SNR per pixel is improved by a factor of 2 (M2 and M5) and $\sqrt{10}\simeq$3.162 (M3 and M6). 
In the next three rows we show mock images created with the BELLS1 lens models and a resolution of 0.09 arcsec, but with different source models: $(i)$ a double Gaussian distribution closely following the light distribution of the original BELLS1 source (M7), $(ii)$ a single larger Gaussian model (M8) and $(iii)$ the image of the low-$z$ galaxy NGC 5457 (M9). In the sensitivity maps, the colour scale shows the lowest detectable mass $M_{\rm low}$ in units of $\log M_{\odot}$, when the perturbation is an NFW halo at the redshift of the lens.}
    \label{images_hst}
\end{figure*}

\begin{figure*}
     \centering
    \includegraphics[width=0.75\hsize]{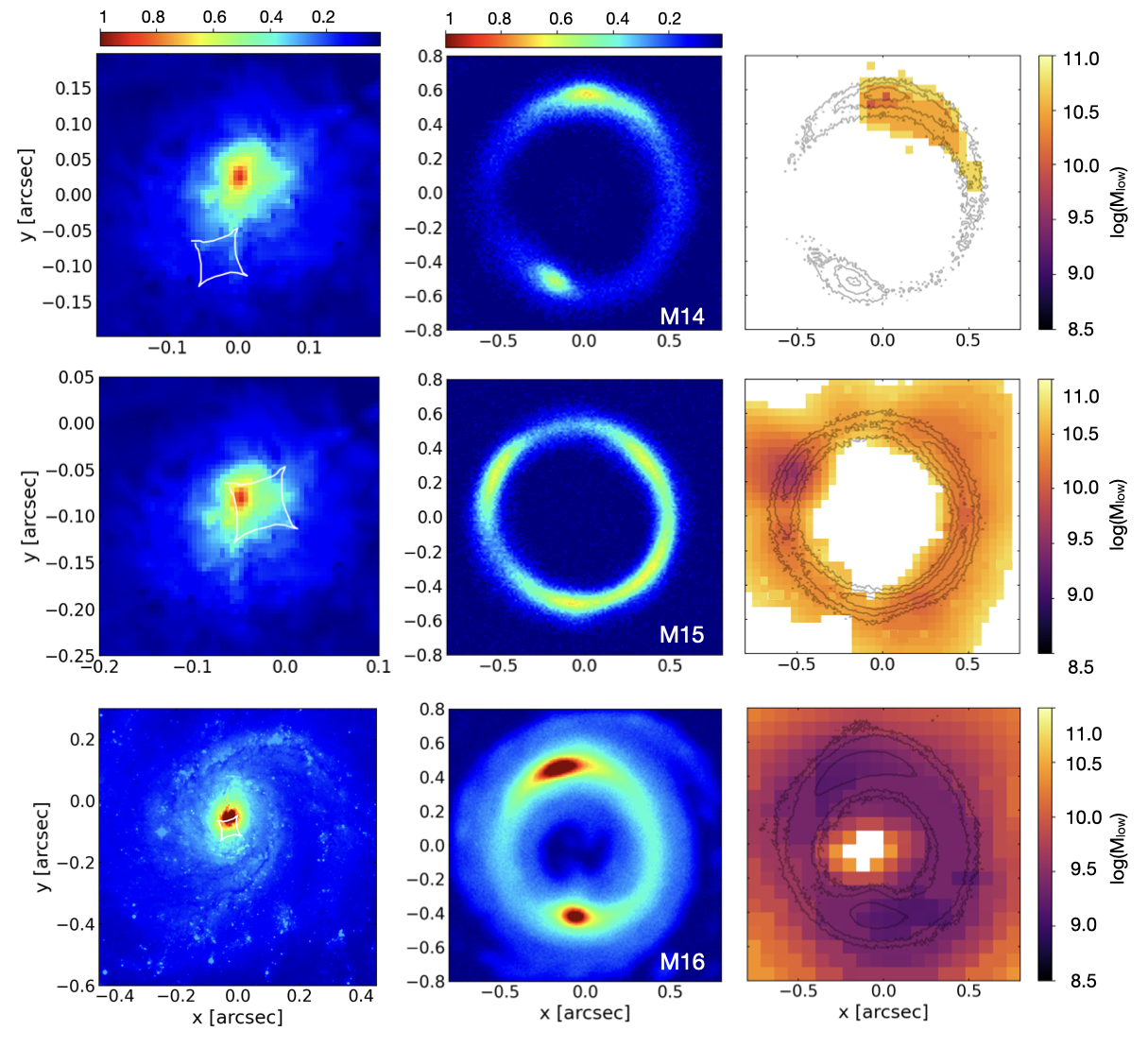}
    
\caption{Summary plot showing the mock data M14 to M16, simulated with an angular resolution of 0.07 arcsec. From left to right we show the source model, the lensed images and the corresponding sensitivity maps for each case. The caustics lines are shown in white on the source plane. The source used for M14 and M15 is the best reconstructed model for the lens system SHARP1, while for M16 we used the image of the low-$z$ galaxy NGC 5457 - see Table \ref{tab_mocks}. The colour scale of the sensitivity maps shows the lowest detectable mass in each pixel, $\log M_{\rm low}[M_{\odot}]$, as in Figure \ref{images_hst}.}
    \label{images_ao}
\end{figure*}

\begin{figure*}
     \centering
    \includegraphics[width=0.95\hsize]{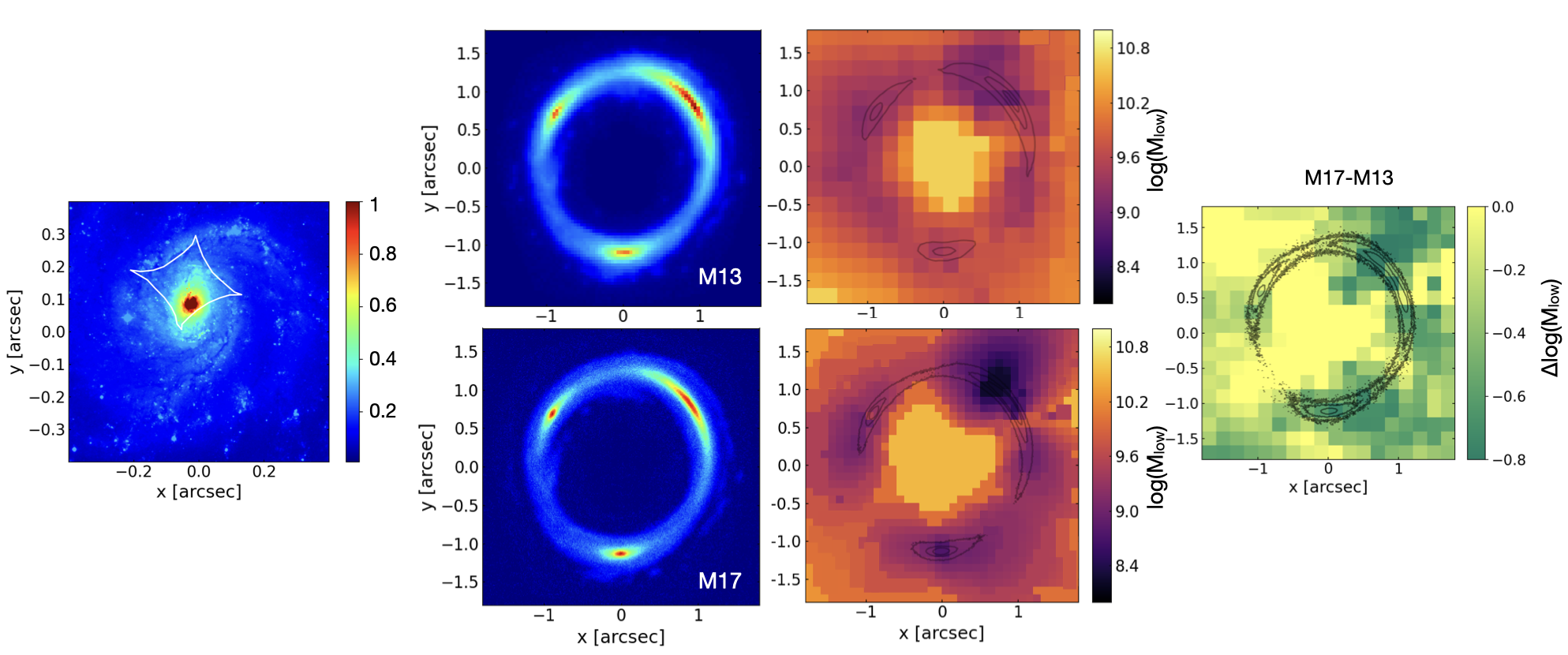}
    \caption{Comparison between images and sensitivity maps with different angular resolution. From left to right we show: $(i)$ the source used in both cases, $(ii)$ the lensed images at 0.09 (top) and 0.07 (bottom) arcsec resolution and $(iii)$ the corresponding sensitivity maps. Finally, the last panel shows the difference in sensitivity calculated as $\Delta\log(M_{\rm low})=\log M_{\rm low}(M17)-\log M_{\rm low}(M13)$. In these mock images, we matched the SNR per unit area, so that the only difference between the two images lies in the angular resolution and pixel size.}
    \label{images_hstao}
\end{figure*}

The mock images are summarised in Table \ref{tab_mocks} and (some are) shown in Figures \ref{images_hst}, \ref{images_ao} and \ref{images_hstao}.

\section{Sensitivity for improved data}
\label{results}

We now discuss the analysis of the sensitivity maps, calculated as described in  Section \ref{method_sens} and shown in Figures \ref{images_hst}, \ref{images_ao} and \ref{images_hstao}. In Figure \ref{exp_all}, we summarise the results for all  systems, in terms of the sensitive area and the minimum and mean values of $M_{\rm low}$ in the maps.

In particular, in this section we describe our findings for mocks with increasing SNR and angular resolution. The simulated data sets M10, M11, M12, M18 and M19 turned out to be completely  insensitive to NFW haloes in the explored mass range and thus they are not shown in the figures but only discussed in the text.

\subsection{The effect of the signal-to-noise ratio}

We use the mock images created with the same lens and source properties but varying signal-to-noise ratio, to investigate how the latter affects the sensitivity to low mass haloes. For this, we use two sets of systems M1-M2-M3  and M4-M5-M6: in both cases, the SNR is progressively increased from the original value in the observations (SNR of $\simeq4$ in M1 and M4), by a factor of $\sim$ 2 and $\sim$3.16. The angular resolution is 0.09 arcsec and the pixel scale is 0.04 arcsec in all images; the size of the lens plane is (3.6 arcsec)$^2$, corresponding to $\simeq$(27 kpc)$^2$ at z=0.73 (BELLS1) and $\simeq$(24 kpc)$^2$ at z=0.56 (BELLS2).

The sensitivity maps are summarised in Figure \ref{images_hst} (rows 1 and 2). For each system, we show: the source, the mock images at the reference SNR (M1 and M4) and the sensitivity maps for the three different levels of SNR. The colour-scale represents the minimum detectable mass at the redshift of the lens ($M_{\rm low}$) at each location, expressed in units of $\log(M_{\odot})$. We immediately see that an increase in SNR generally improves the sensitivity. The effect is twofold: the value of $M_{\rm low}$ consistently decreases at each considered location and, as a consequence, the sensitive area also becomes larger with increasing SNR. On average, $M_{\rm low}$ decreases linearly with SNR as: 

\begin{equation}
    \Delta\log M_{\rm low} = \log \frac{M_{\rm low}}{M_{\rm low,0}} = 1.5(\pm0.1) -0.725(\pm0.12)\times \mathrm{\frac{SNR}{SNR_{0}}},
    \label{eq_snr}
\end{equation}
where $M_{\rm low,0}$ is measured with the original SNR level (i.e. SNR$_{0}$).

A higher SNR significantly extends the sensitive region far from the lensed arc: for M1 and M4, only the brightest lensed images are sensitive to masses lower than $10^{10}M_{\odot}$, while this is not the case any more at higher SNR. The left panel in Figure \ref{mlow_dep} shows the (logarithmic) decrease in $M_{\rm low}$ for the two levels of improved SNR (green and purple contours and point) as a function of the value of $M_{\rm low}$ at the original SNR level. The detectable mass improves everywhere (i.e. decreases) and the improvement is larger at the high-mass end: a large number of pixels with very little sensitivity (i.e. high $M_{\rm low}$) at SNR$_{0}$ becomes sensitive thanks to the higher SNR. The top part of Figure \ref{exp_all} shows, for all lenses, the area of the lens plane where haloes of mass $M>M_{\rm low}$ can be detected (in arcsec$^2$), clearly demonstrating the increase with SNR for all values of $M_{\rm low}$. Of course, we do not expect $M_{\rm low}$ to improve indefinitely with SNR: the image resolution puts a relatively hard lower limit on the size of the smallest detectable perturbation. The highest SNR considered here corresponds to an increase in observational time of a factor of 10 and thus already much higher than the currently available real data sets: we thus consider it to be a realistic upper limit. Moreover, given that the SNR is proportional to the square root of observational time and that the angular resolution ultimately sets the size of the smallest detectable perturbation, we expect the sensitivity to reach a resolution floor as observation time increases further.

We created all the mock images presented here from the reconstructed sources of real observations with the lowest SNR considered. As a result our higher SNR images may be missing small scale features compared to an actual observation of comparable quality. For this reason, our result should be interpreted as conservative.

\subsection{The effect of the angular resolution} 
\label{sec_ao}

We now investigate the effect of the angular resolution by comparing the simulated data sets M14 to M17 to their lower-resolution counterparts, systems M10 to M13. Because the set of source and lens galaxies is the same, we can perform a one-to-one comparison between each pair of images. The group M14-M17 has a PSF FWHM of 0.07 arcsec and a pixel scale of 0.01 - as the original observations from which the model is taken (see Tables \ref{tab_obs} and \ref{tab_mocks}) - while the set M10-M13 has been created with a PSF FWHM of 0.09 arcsec and a pixel scale of 0.04, as in the previous section. As described in Section \ref{sec:noise_models}, each case we match the SNR per unit area across the two images. 

\begin{figure*}
     \centering
    \includegraphics[width=\hsize]{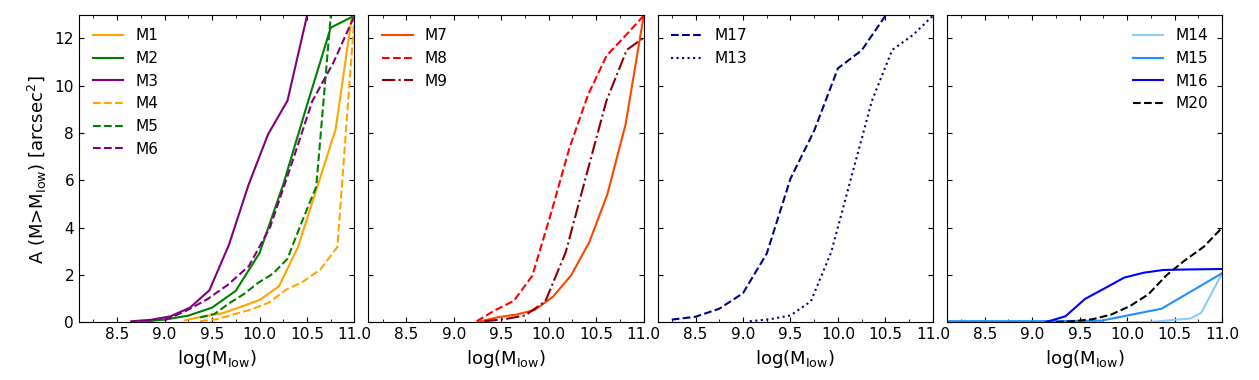}
    \includegraphics[width=0.75\hsize]{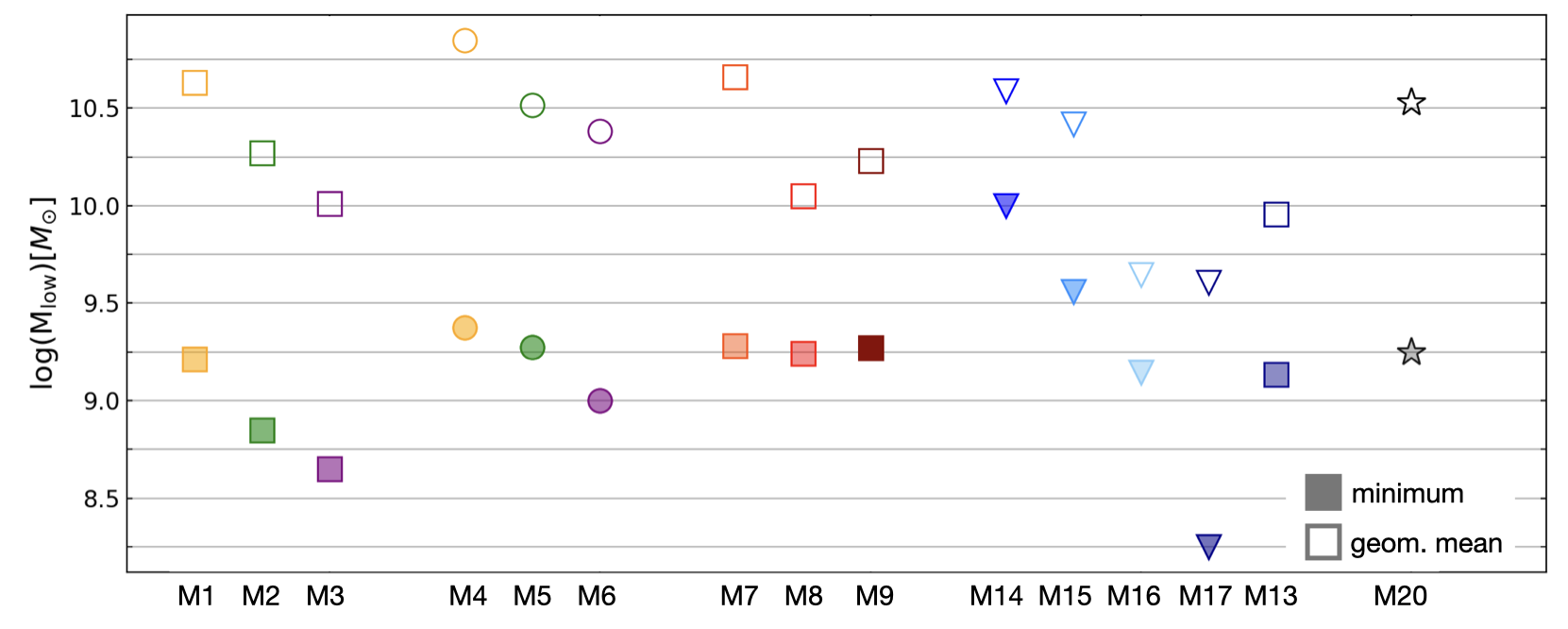}
    \caption{\emph{Top}:for each lens, we show the area (in  arcsec$^2$) on the lens plane, where haloes of mass $M>M_{\rm low}$ can be detected. From left to right, we summarise results $(i)$ at different SNR, $(ii)$ source structures, $(iii)$ higher vs. lower resolution and $(iv)$ for the Keck-like and the ALMA mock data. \emph{Bottom}: for each system, we show the minimum (filled symbols) and the mean (empty symbols) value of $M_{\rm low}$ in the corresponding sensitivity map.
    }
    \label{exp_all}
\end{figure*}

Our first finding is that the images M10-M12 are not sensitive to NFW haloes in the considered mass range - this lack of sensitivity results from the low resolution combined with the small size of the Einstein radius, as discussed in the Section \ref{configuration}.  Conversely, their counterparts M14-M16 are sensitive and we show the sensitivity maps in Figure \ref{images_ao}. This demonstrates that indeed, a small increase in data resolution can have a significant effect in terms of sensitivity: potentially, re-observing promising systems at a higher angular resolution could be an effective observational strategy to increase the number of expected detections. The $M_{\rm low}$ range in Figure \ref{images_ao} is similar to that of Figure \ref{images_hst}; the sensitive area in physical units (top-right panel of Figure \ref{exp_all}) is however smaller, due to the small size of the Einstein radius.
\citet{vegetti12} reported a low-mass detection in the system B1938+666 (corresponding to our mock system M14), at the location of the brightest part of the arc. This detection might seem not entirely compatible with the sensitivity map presented in Figure \ref{images_ao}. However, we remind the reader that here we model perturbations as NFW profiles, while \citet{vegetti12} detected this subhalo in a model-independent pixellated fashion and showed it to be consistent with a Pseudo-Jaffe (PJ) profile. When we consider a PJ profile, we find values of $M_{\rm low}$ consistent with the mass detected in the real data. Our findings, together with the recent work by \citet{minor21a}, point to the fact that not all dark sources of perturbations can be well described by NFW profiles. We will investigate this in a follow-up paper.

\begin{figure*}
    \centering
    \includegraphics[height=0.38\hsize]{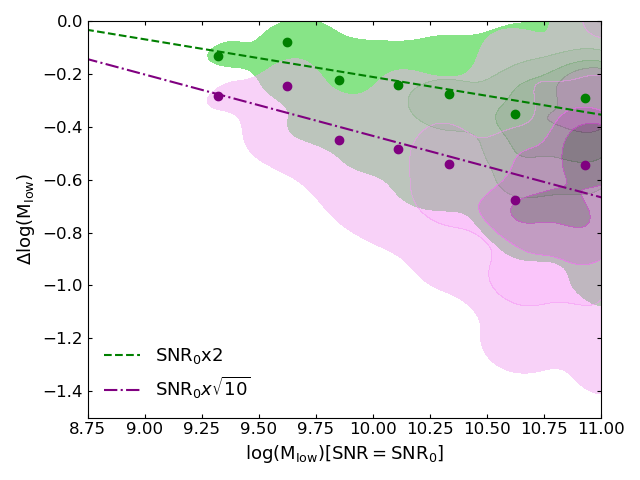}
    \includegraphics[height=0.38\hsize]{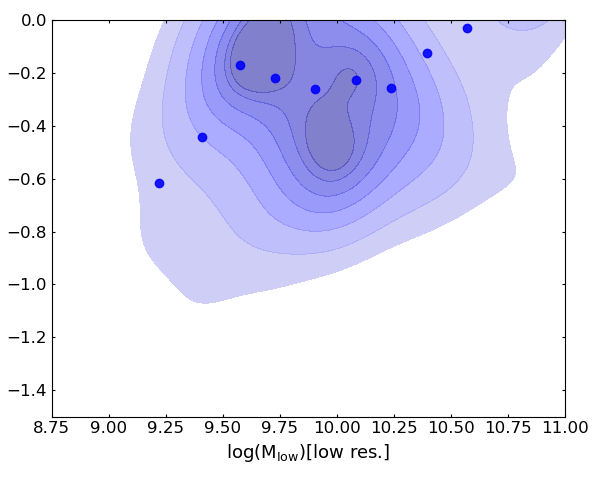}
    \caption{Improvement in sensitivity with SNR (left) and angular resolution (right). In both cases, we show the logarithmic difference in $M_{\rm low}$ expressed as $\Delta\log(M_{\rm low}) = \log(M_{\rm low}/M_{\rm low,0})$. The coloured contours show the density of points in the considered plane: in the left panel the improvement due to the increased SNR with respect to the reference one SNR$_{0}$ (see Figure \ref{images_hst}) and in the right panel the improvement due to the higher resolution (0.07 on the $y$-axis vs 0.09 arcsec on the $x$-axis - see Figure \ref{images_hstao}). The points stand for the mean values as a function of $\log(M_{\rm low})$ and the lines in the left panel correspond to the best-fit linear relation to the points.}
    \label{mlow_dep}
\end{figure*}

In Figure \ref{images_hstao} we show the results for M13 and M17: in this case both images have a good level of sensitivity and can be directly compared. The rightmost panel shows the improvement in the sensitivity (expressed as $\Delta\log M_{\rm low}$) at each considered location on the lens plane. In this case, the average gain in $M_{\rm low}$ is 0.25 dex at all masses, with a larger spread at the intermediate values of $M_{\rm low}$ and a tendency for larger improvements at the locations where the sensitivity is already good (see the right panel of Figure \ref{mlow_dep}). The effect of a higher angular resolution is thus different from that of a higher SNR: the largest improvement (and spread in $\Delta\log M_{\rm low}$) in this case is seen in the most sensitive pixels and not at high masses, opposite to the SNR effect shown in the left panel. The change in the sensitive area as a function of $M_{\rm low}$ is shown in the third panel in the top row of Figure \ref{exp_all}; the effect on the most sensitive region can be appreciated by comparing the mean and minimum values of $M_{\rm low}$ in the bottom panel of Figure \ref{exp_all}. This is a promising result, but it should be interpreted with caution, given that it is based on only one system and the source model used here is a rescaled low-$z$ spiral galaxy, that could be unrealistic in terms of morphology.

\begin{figure*}
     \centering
    \includegraphics[width=0.9\hsize]{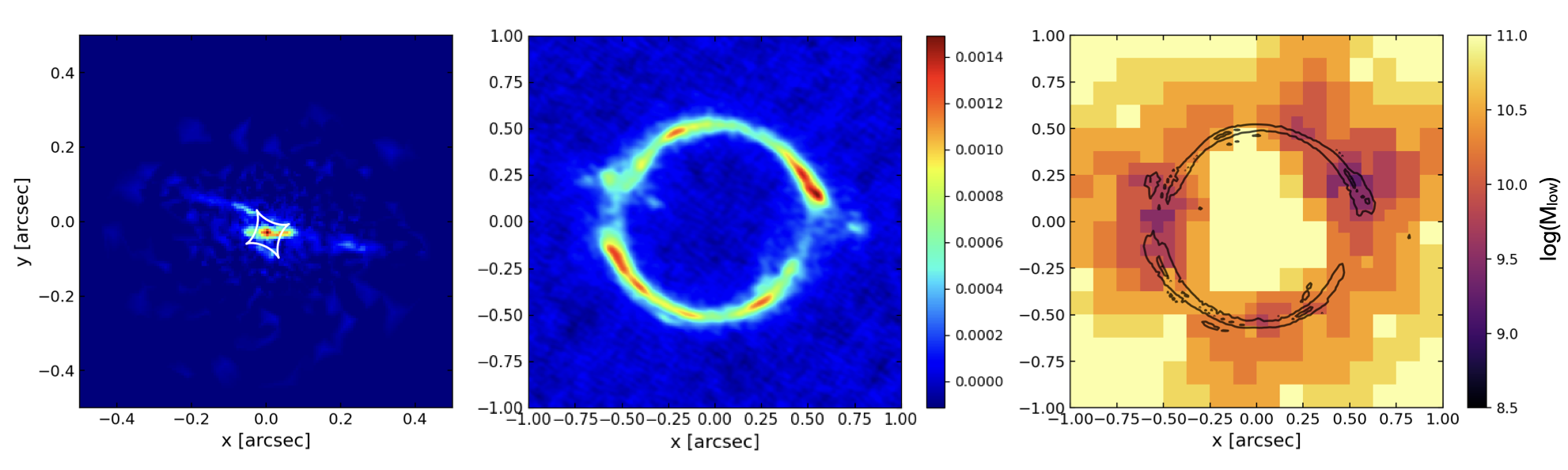}
    \caption{Source (with caustics), deconvolved image and sensitivity map of the system ALMA1 (SPT\,0532$-$50). The sensitivity is refined with a grid scale of 0.0625~arcsec near where there is lensed emission.}
    \label{alma1}
\end{figure*}

Finally, we calculate the sensitivity maps for the two systems with the lowest resolution M18 and M19 (PSF FWHM of 0.16 arcsec, pixel scale of 0.1 arcsec) and found them insensitive to NFW haloes in the mass range $M\leq5\times10^{10}M_{\odot}$. This result may have important implications for future observations of comparable angular resolution and SNR, such as those with Euclid. We discuss this further in Section \ref{other_data}.

\subsection{Interferometric observations with ALMA}
Here we explore the potential of detecting low-mass haloes with ALMA, by analysing one simulated image. ALMA1 as a simulated ALMA observation of a gravitationally lensed dusty star-forming galaxy \citep[for example,][]{negrello10,alma_dusty,dsb_galaxies}. To create this mock data, we use band 7 observations of SPT\,0532$-$50 at 0.045~arcsec resolution (project code 2016.1.01374.S; PI: Hezaveh). We chose this system because it provides the best combination of SNR, angular resolution, and source redshift of currently available data in the ALMA archive. The real data were modelled using the pixellated method of \cite{vegetti09} extended for use with interferometric data as described in \cite{powell21}. Further details of the data reduction and lens modelling will be given in Stacey et al. (in prep). The mock data were created by directly overwriting the visibility data with the Fourier-transformed sky model, then adding Gaussian noise at the same level as measured in the actual data. From the sensitivity map calculation, we find these observations to be sensitive to haloes at the redshift of the lens in the mass range $M\geq1.8\times10^{9}M_{\odot}$. We plot the source, images and sensitivity map in Figure \ref{alma1}.

The range of $M_{\rm low}$ in Figure \ref{alma1} is similar to that of the Keck-like data in Figure \ref{images_ao}. In the bottom panel of Figure \ref{exp_all}, the minimum and mean value of detectable mass $M_{\rm low}$ are represented by the black stars: while the mean value is quite high, the minimum is comparable to the best Keck-like realistic data, thanks to the high angular resolution.

\section{The impact of the lensing configuration}
\label{configuration}

In addition to angular resolution and SNR, several other factors - specific to each observed system - can influence the number of potential detections. Here we test the most important ones in a controlled set-up. We refer to the sensitivity maps in Figure \ref{images_hst} and \ref{images_ao}, while indicators for all systems are summarised in Figure \ref{exp_all}.

\subsection{Source structure}

\citet{blanford01} and \citet{koopmans05} have shown that the sensitivity to low mass haloes is directly related to the gradient of the source galaxy light - the larger the complexity of the source light, the lower the detectable mass \citep[see also][]{rau13}. This section systematically quantifies this effect by comparing the sensitivity map of four simulated images characterised by different source properties: 

\begin{itemize}
    \item[--] in M1 the source  is a compact ($\sim$1.5 kpc) clumpy galaxy, typical of Lyman-$\alpha$ emitters at high-redshift - this corresponds to the original source in the HST data;
    \item[--] in M7 the source light distribution is described by a double Gaussian, created in order to closely reproduce the shape of the source in M1;
    \item[--] M8 has single Gaussian source that is larger (in size) and brighter than M1;
    \item[--] in M9 the source is the real galaxy NGC5 457, properly rescaled in size.
\end{itemize}

All the sources are placed at the same location relative to the caustic curves, and their source brightness is rescaled to span the same total range as in M1; however, due to the different light structures, the overall distribution of SNR can be different. In all cases, the PSF FWHM is 0.09 arcsec, and the pixel scale is 0.04 arcsec. The lens and source redshift are also the same as in M1 (see Table \ref{tab_mocks}). The sources, lensed images and sensitivity maps are shown in Figure \ref{images_hst}.

We find that the sensitivity of M7 is very similar to that of M1, in agreement with the fact that they have very similar source properties. However, the M1 source has a more complex distribution and more extended low surface-brightness tail, resulting in a sensitivity more extended far from the brightest lensed images.
In M8, the source has by definition no small scale granularity; however, the more extended bright peak results in a more extended brighter arc and lower values of $M_{\rm low}$. An intermediate level of improvement is produced by the more extended - but with a smaller central peak - source in the last panel. Thus we confirm that a more structured and brighter source plays an important role in the sensitivity and, as a consequence, in the number of detectable (sub)haloes. This is shown more quantitatively in Figure \ref{exp_all}: in the second panel of the top row we can see the change in the sensitive area (the largest in M7, and intermediate in M8) and, in the bottom panel the difference in the mean value of $M_{\rm low}$.

\subsection{Source position}

The position of the source with respect to the caustic lines determines which fraction of the source surface brightness is highly magnified and this has an impact on the data sensitivity. We demonstrate this with two examples: 

\begin{itemize}
    \item[--] the source used for M1-M3 has a larger fraction inside the caustic lines relative to the source in M4-M6 (see Figure \ref{images_hst}) and, as a result, the first set has a higher sensitivity: both the mean and the minimum values of $M_{\rm low}$ are lower and the sensitive area is larger (see Figure \ref{exp_all}).
    \item[--] We see this further by comparing M14 with M15 (Figure \ref{images_ao}), where the same source model is placed at two different locations: in M15 a centrally-aligned source enhances the magnification of its brightest part, creating more extended arcs and thus a larger sensitive area. Even if the overall median SNR is comparable, M15 is clearly a more promising system for dark matter studies: in the bottom panel of Figure \ref{exp_all}, we see that the minimum $M_{\rm low}$ value is $\sim$0.5 dex lower than in M14. 
\end{itemize} 

This result shows how preferring systems where the source lies within the caustic lines, and it is thus more magnified, is a good strategy to maximise possible detections, due to the increased sensitive area.

\subsection{Size of the Einstein radius and redshift configuration}

The total number of expected detections can differ significantly from one lens system to another, even when the range of detectable masses is similar. The size of the Einstein radius (see Table \ref{tab_mocks}) determines the maximum extent of sensitive area on the lens plane, used for integrating the (sub)halo mass function. Moreover, for the case of field haloes, the redshift of the lens and source set the size of the cosmological volume probed by each lens system \citep{despali18}.

Qualitatively, systems with a larger Einstein radius, a higher lens and source redshift lead to a higher number of detections for a fixed SNR, angular resolution and source structure. More quantitatively, we can gauge the magnitude of this effect by comparing the simulated data sets M17 and M16. These lens systems have the same source but a different Einstein radius due to the different lens models used to create them. The systems have similar angular resolution; the mean detectable mass (see Figure \ref{exp_all}) is similar in both cases - while the minimum of $M_{\rm low}$ differs quite significantly. The number of detectable objects will also depend on the size of the Einstein radius (1.4 vs 0.45 arcsec) and the redshift of the source (2.5-2.13 vs 2.01) - when including the contribution from field haloes. They both lead to a larger volume and thus a higher number of potential detections.
Thus, high angular resolution observations are the most promising to extend the mass range of potential detections to low masses, but ideally, this has to be complemented by a larger cosmological volume.

\section{Perspective on future observations} \label{other_data}

Strong gravitational lensing is a relatively rare phenomenon, and current samples of known lenses amount to roughly a few 100s. Of these systems, however, not all of them have the necessary data quality to provide stringent constraints on the properties of dark matter. For example, the SLACS sample  \citep{bolton06} has  a total of almost 100 lens systems; however,  \citet{vegetti14} selected the systems with the best SNR (11 lenses) and modeled them to search for the presence of low-mass haloes. A similar result was obtained by \citet{ritondale19a} who showed that only 14 systems of the 17 that they analysed could provide some sort of constraints on the halo mass function. The SPT and Herschel-ATLAS surveys have discovered a total of more than one hundred lenses, however until now only a few have follow-up ALMA observations \citep[for example 26 from][]{dsb_galaxies} and among these a small subset ($\sim3-4$) has a resolution of $\leq0.025$ arcsec \citep{hezaveh13}. There are currently only a handful of data sets with angular resolution below 0.05 arcsec that have comparable surface brightness sensitivity to the mock data studied here.

In this context, our results help to understand the system-to-system variation of the actual data sensitivity and to derive realistic predictions. Figure \ref{exp_all} provides a summary of the results discussed so far: the top panel shows the area of the lens plane that is sensitive - and thus relevant for dark matter detection - for each system, while the bottom panel summarises the minimum and mean values of the lowest detectable mass $M_{\rm low}$ over the lens plane. While we have modified one property at a time in a controlled fashion, in reality a combination of the effects described in the previous sections can apply to each new observation. Overall, the best chances of detecting a low mass (sub)halo are given by a very high SNR (purple points, M3 or M6), possibly combined with a high angular resolution and bright source (blue triangle, M17). Moreover, one would ideally observe systems with a large Einstein radius. While the mean $M_{\rm low}$ spanned by the systems M14-M16 is comparable, if not better, to that of M1-M6, the total sensitive area is smaller - see the upper panel of Figure \ref{exp_all}. This is the result of the initial selection of the SHARP lenses: they were not lens-selected (i.e. targeting massive elliptical galaxies as possible lenses) in the same way as the BELLS and SLACS surveys, but rather were taken from source-selected searches (CLASH and SPT) and as a result the sample contains both high- and low-mass galaxies. Keck observations of lenses with larger Einstein radii will provide higher angular resolution which will be important for detecting low mass haloes and distinguishing CDM from alternative models. 
  
Despite the differences in the observational set-up and resolution of the systems considered here, the smallest value of the lowest detectable mass $M_{\rm low}$ is between $3\times10^{8}$ and $3\times10^{9}M_{\odot}$ in most cases (see the bottom panel of Figure \ref{exp_all}). Only in one best-case scenario, the lower limit reaches $1.6\times10^{8}M_{\odot}$. This shows how new even higher resolution observations will be essential to push the limit further down in mass: we now discuss in more details what are the best near-future opportunities to achieve this goal.
  
The Keck telescope will undergo a significant upgrade soon, with the Keck All-sky Precision Adaptive Optics (KAPA). This upgrade will be beneficial in a couple of ways: (i) it will allow us to cover a more significant fraction of the sky and, therefore, expand the current samples of Keck-observable lenses; (ii) the AO PSF will improve in terms of the Strehl ratio, increasing the SNR at fixed integration time. With a resolution of a few mas, the upcoming extremely large telescopes (i.e the ELT, GMT and TMT)
will be a game-changer, allowing us to set tight constraints on the halo mass function with a relatively limited number of lens systems, provided that a good model for the PSF is available (Vegetti et al. in prep.). 

At cm- and sub-mm-frequencies, ALMA and especially VLBI observations already provide a viable and successful strategy to reach the interesting angular resolutions of 0.025-0.075 arcsec \citep{alma15} and milli-arcsec \citep{mckean15}, respectively. In the case of ALMA, the angular resolution is comparable with the most sensitive data considered in this paper. However, one has to consider the required observing time: for the ALMA data considered in this work (M20), the integration time was 4 hours, thus reaching a number of expected detections significantly above one might prove to be extremely challenging. One approach would be to target a small number of promising candidates, re-observing them with a longer integration time and with the longest baseline, to reach 30 mas resolution.  As proven by the system considered in this work, ALMA can be used to target high redshift objects, thus maximising the combined effect of high angular resolution and probed cosmological volume.

Shortly,  imaging and interferometric surveys carried out by the LSST (0.2-0.7 arcsec resolution), MeerKAT (> 1 arcsec) and the SKA (sub-arcsec resolution at high radio frequencies, e.g. 0.05 arcsec at 10 GHz) will together lead to the discovery of several thousand new strong gravitational lens systems; Euclid (0.16-0.3 arcsec resolution) alone is expected to discover over 10$^5$ new lensed galaxies \citep{Euclid_report}. However, as demonstrated in this paper, the angular resolution of these instruments might not be sufficient to detect objects at the low-mass end of the (sub)halo mass function (when perturbers are modeled with NFW profiles) and thus obtain strong statistical constraints on CDM and alternative WDM models. If one allowed for a scatter around the mean concentration-mass relation and thus some perturbing (sub)haloes were more concentrated than the standard NFW profile considered here, it could in principle be possible to obtain more optimistic forecasts \citep{amorisco21}. At the same time, according to \citep{minor21a}, more detailed studies might be necessary to clarify the nature of the detected perturbing objects and their properties to check if they are indeed well modeled by NFW profiles of any concentration. In both cases, high-resolution follow-up observations will be advisable and would allow one to obtain more stringent constraints. The design of an effective follow-up strategy for the Euclid sample will have to consider more than one critical aspect: we have shown in Section \ref{results} that the angular resolution of the observations is not the only important parameter, with the redshift configuration of the system and the properties of the lensed sources also playing key roles. In this respect, the lens and source redshift distribution of the gravitational lens systems that Euclid will discover will likely peak around 0.5 and 2.0 arcsec, respectively.  In terms of detecting field haloes, this redshift configuration is better than SLACS but worse than the BELLS-GALLERY sample. Most of the  Euclid systems will have an Einstein radius around 0.5 arcsec, resulting in a potential loss of sensitive volume compared to the SLACS sample. Moreover, the source galaxies are also expected to be fainter, potentially requiring longer integration times.

Finally, the JWST could, in principle, be an ideal instrument to target high-redshift lensed galaxies with a moderate angular resolution (0.06-0.1 arcsec). In practice, however, this strategy could prove somewhat challenging, because the best angular resolution will be achieved in the bluest filters, where high-$z$ galaxies might not be bright enough to be observed without a significant investment of telescope time.

\section{Conclusions} 
\label{concl}

Strong gravitational lensing is one the most robust probes of the halo mass function and the nature of dark matter. As we enter a new golden era for this field, the large sample of lenses soon to be discovered coupled with high-resolution follow-up observations represent a unique opportunity to improve our knowledge of the dark sector. However, follow-up observations will likely only be possible for a subset of the several thousand new objects. Developing an observing strategy to maximise the number of detectable low-mass haloes and the statistical strength of non-detections is critical to fully taking advantage of this opportunity.

This work is the first step towards a systematic understanding of all the different factors that make a specific strong lensing system and relative observations more or less suitable to test the CDM paradigm and alternative WDM models. Using simulated observations, we have found that a number of factors play a role in setting the mass sensitivity of the data, demonstrating the complexity and the challenging nature of constraining the properties of dark matter with strong gravitational lensing: the SNR, the brightness and position of the source, the angular resolution and the observed wavelength. For this reason, one has to be extremely careful with constraints and predictions which have not been tuned to real systems and that consider a constant value of $M_{\rm low}$ over the entire lens plane. We find that the main factors are:
\begin{itemize}

    \item the signal-to-noise ratio of the data: the sensitivity improves as a consequence of a higher SNR (or longer observing time) and the lowest detectable mass scales with the SNR following equation (\ref{eq_snr}). An increase in SNR can significantly extend the sensitive region on the lens plane, beyond the location of the brightest lensed images;
    \item the angular resolution of the data, which determines the minimum mass of the detectable objects. When the angular resolution improves even mildly from 0.09 to 0.07 arcsec, the lowest detectable mass on the lens plane decreases on average of 0.25 dex in log(M), reaching $1.7\times10^{8}M_{\odot}$; the effect on the distribution of detectable masses (for a fixed lensing configuration) is comparable to an increase in observational time by a factor of $\sim4$. Moreover, an increase in angular resolution drives an improvement in sensitivity especially in the regions sensitive to the lowest masses, i.e. the minimum value of $M_{\rm low}$;
    \item the size of the Einstein radius and thus of the area on the sky relevant for the detections;
    \item the structure and position of the source galaxy: a brighter and more structured source and a source located mostly within the caustic lines (thus more magnified) can lead to an improvement in the mean $M_{\rm low}$ of $\sim$0.5 dex.
\end{itemize}

We conclude, therefore, that, in the ideal scenario where one had complete control on these parameters, one should target high-redshift systems, where a bright source is lensed to extended arcs, with highest possible angular resolution. 
Moreover, it is worth requesting longer observation runs to improve the SNR of the data - even at fixed resolution, which has a significant impact on the number of detectable objects. 

The key ingredient of this analysis is the calculation of a detailed sensitivity map for each system (described in Section \ref{method}), that measures the lowest mass $M_{\rm low}$ that can be detected given the data at a certain level of significance. We have demonstrated that such maps are an essential tool and that predictions calculated from idealised mock data or assuming a constant value of sensitivity could produce overoptimistic predictions. 

In this paper, we have focused on how the observational setup influences the data sensitivity. In the second paper of this series, we will derive the number of \emph{field} haloes that can be detected by each configuration and derive predicted constraints on CDM and WDM models. It is also necessary to investigate further the effect of profiles and concentrations deviating from the main scaling law for field CDM haloes, given that this could have a substantial impact on the final sensitivity, as demonstrated in Appendix \ref{macromodel} and in \citet{amorisco21}.  
It is fair to discuss limitations also on the theoretical side: numerical simulations with alternative dark matter model have not yet reached the same level of complexity of standard CDM hydrodynamical simulations and their number and size is limited. Moreover, as found by \citet{minor21b}, the structure of at least one observed subhalo appears inconsistent with that of typical subhaloes in current hydrodynamical galaxy formation simulations. For this reason, future simulations will improve our understanding of the interaction between alternative dark matter and baryonic physics and provide more robust measurements of the halo and subhalo mass functions at small scales.

\section*{Acknowledgements}
GD thanks Carlo Giocoli, Matt Auger and Conor O'Riordan for useful comments and discussions. We thank the referee, Simon Birrer, for his insightful comments on the paper. SV thanks the Max Planck Society for support through a Max Planck Lise Meitner Group; SV, DP and HS acknowledge funding from the European Research Council (ERC) under the European Union’s Horizon 2020 research and innovation programme (LEDA: grant agreement No 758853). This research was carried out on the High Performance Computing resources of the FREYA and COBRA clusters at the Max Planck Computing and Data Facility (MPCDF) in Garching operated by the Max Planck Society (MPG).
CDF acknowledges support from the National Science Foundation  under Grant No. AST-1715611.

\section*{Public data and software}
The work presented in this paper is based on observations made with the W. M Keck Observatory and the NASA/ESA Hubble Space Telescope. Observational data were obtained from the Data Archive at the Space Telescope Science Institute, which is operated by the Association of Universities for Research in Astronomy, Inc., under NASA contract NAS 5-26555. These observations are associated with program 14189. We made use of ALMA data with project code 2016.1.01374.S. ALMA is a partnership of ESO (representing its member states), NSF (USA) and NINS (Japan), together with NRC (Canada), MOST and ASIAA (Taiwan), and KASI (Republic of Korea), in cooperation with the Republic of Chile. The Joint ALMA Observatory is operated by ESO, AUI/NRAO and NAOJ. This research made use of Astropy, a community-developed core Python package for Astronomy \citep{astropy13}, of the matplotlib \citep{matplotlib}, SciPy \citep{scipy} and NumPy \citep{numpy} packages.

\section*{Data availability} 
The images and mock data analysed in this work are available upon request to the corresponding author. Moreover, the original HST, Keck-AO and ALMA images for the considered systems are publicly available in the corresponding archives.




\bibliographystyle{mnras}
\bibliography{mnras_template.bbl} 



\appendix

\section{Main lensing potential} \label{app_mm}

From an analysis of the SLACS lenses assumint a Pseudo-Jaffe (PJ) mass density profile for the subhaloes, \citet{vegetti14} found that the degeneracy of the subhalo parameters with the source structure was larger than the degeneracy with the mass density profile of the host lens. For this reason, they calculated the sensitivity function for their sample by keeping the macro-model parameters fixed at their most probable value and only marginalised over the source light and its regularisation level.

\citet{despali18} have shown that NFW subhaloes are more diffuse than PJs because their lensing effect is weaker and less localised. Here, we quantify the degeneracy between an NFW profile and the macro-model parameters and its effect on the sensitivity function calculation. 

\begin{figure*}
     \centering
    \includegraphics[width=0.8\hsize]{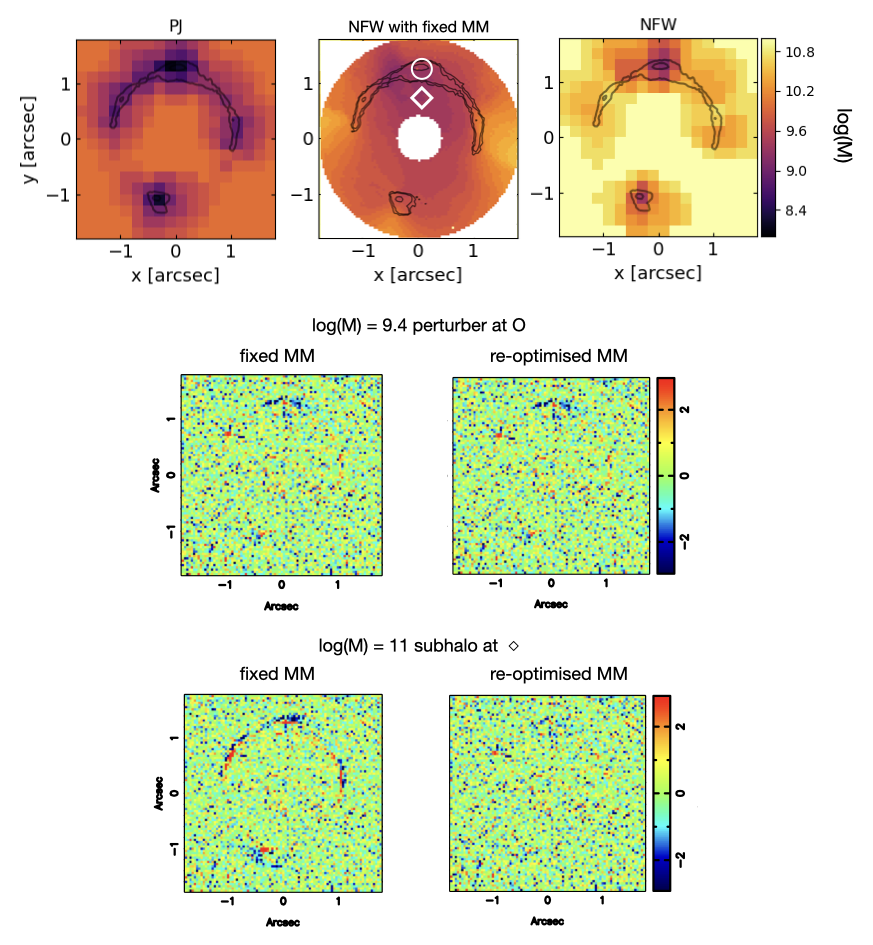}
    \caption{Example of the effect of using a fixed macro model during the sensitivity calculation vs. re-optimising the main lens parameters at the same. The sensitivity maps with fixed or marginalised macro model (MM) are shown in the top-middle and top-right panels. The top-left panel shows instead the sensitivity map for a Pseudo-Jaffe profile, that is not affected by keeping the macro model fixed as much as the NFW case. Note that the difference range in the values of $M_{\rm low}$ is due to the fact that the enclosed mass is defined differently for the two profiles, as detailed in \citet{despali18}. The other two sets of panels show the residual between the data (not including the perturber) and the best model (including the pertuber), both for the case where the main lens model (i.e. the macro model) is kept fixed and the case where it is re-optimised together with the source.}
    \label{macromodel}
\end{figure*}

In Figure \ref{macromodel} (top panels), we compare sensitivity functions calculated for an NFW subhalo obeying the field galaxy mass-concentration relation in two cases: (i) with the parameters of the macro-model fixed to the best values in the absence of a perturbation, and (ii) marginalising over the parameters of the macro-model also for each perturbed case. It is evident that the sensitive regions have completely different shapes in the two cases - while the PJ sensitivity is not affected at the same level and, even with fixed macro-model parameters, shows sensitive regions similar to the right panel. The other panels in Figure \ref{macromodel} show two examples to clarify what the origin of the difference is: in these tests we add a subhalo to the main lens model - the mass $\log(M)=9.4$ in the first case and $\log(M)=11$ in the second case. In the first case, where the subhalo is located in the sensitive region around the arc and has a lower mass, the level of residuals created by a fixed or a re-optimised macro model and their Bayes factor (relative to a smooth model) are similar. In the second case, given that we are dealing with a higher mass far from the lensed images, the residuals and the Bayes factor are much higher for a fixed macro model, leading to an overly optimistic sensitivity map.

The same test for subhaloes with a PJ profile leads to a conclusion in agreement with \citet{vegetti14}. For compact subhaloes, the degeneracy with the macro model is weak, even for objects located further away from the lensed images. This is proven by the fact that the shape of the PJ sensitivity map, with or without a fixed macro model is similar to that of the re-optimised NFW (see Figure \ref{macromodel}, top-left panel). In a follow-up paper, we will study the effect of the subhalo mass density profile in a more systematic way.

\section{Noise realisation} \label{app_noise}

\begin{figure*}
     \centering
    \includegraphics[width=\hsize]{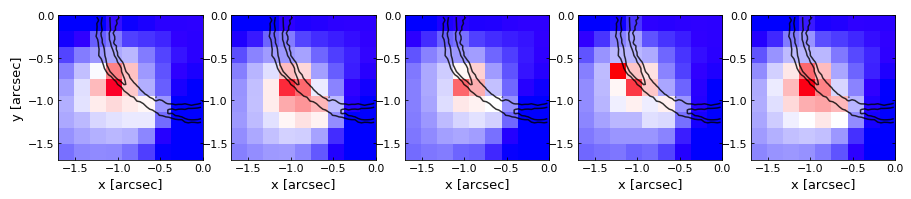}
    \caption{Difference in logarithmic Evidence in five mocks with different noise realisation, calculated for a mass of log(M) = 9.75 in each pixel. For this test, we use focus on the most sensitive quadrant of lens BELLS2. The colour scale ranges from 0 (blue) to 100 (red). This is one of the lowest masses detectable in BELLS2 at $\mathrm{SNR_{0}}$ and it is thus correctly only found ($\Delta\log$Ev>50) in a few pixels along the arc. Even thought the exact values changes, the 
    overall distribution is similar in all panels.}
    \label{sens_noise1}
\end{figure*}

\begin{figure}
     \centering
    \includegraphics[width=0.8\hsize]{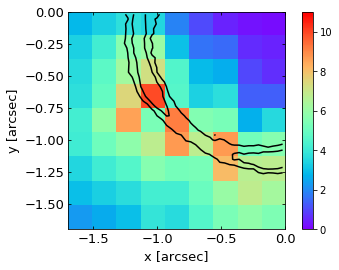}
    \caption{Standard of the difference in logarithmic Evidence for 30 mocks created with different noise realisations.
    }
    \label{sens_noise2}
\end{figure}

Here we discuss the impact of noise on the sensitivity: not SNR, but rather the specific noise realisation present in each mock. For this, we created 30 additional mocks of the system BELLS2 at the original level of noise. We generate a different realisation of Gaussian noise in each mock. Since the sensitivity calculation is expensive, we focused on the most sensitive quadrant (lower-left) of the lens BELLS2. We look at the effect in two ways: $(i)$ we calculate the sensitivity to a low-mass halo of fixed mass in each pixel - where we choose $M=10^{9.75}M_{\odot}$, see Figure \ref{sens_noise1}; $(ii)$ we calculate evidence for the masses of sensitivity map in Figure \ref{images_hst} (see Figure \ref{sens_noise2}), to check by how much they would differ with a different noise realisation.

We find that the measurement of the sensitivity map is quite stable for different noise realisations, leading to fluctuations in evidence of the order of $\simeq$10 and, in turn, to uncertainties in the $M_{\rm low}$ values within 0.15 dex. Although our tests were run only for $\mathrm{SNR_{0}}$, we expect the effect on higher SNR images to be similar. However, in real data sets, where the noise is not known a priori, this is definitely a source of uncertainty.

\section{On the sensitivity threshold} \label{sens_lower}

\begin{figure}
     \centering
    \includegraphics[width=\hsize]{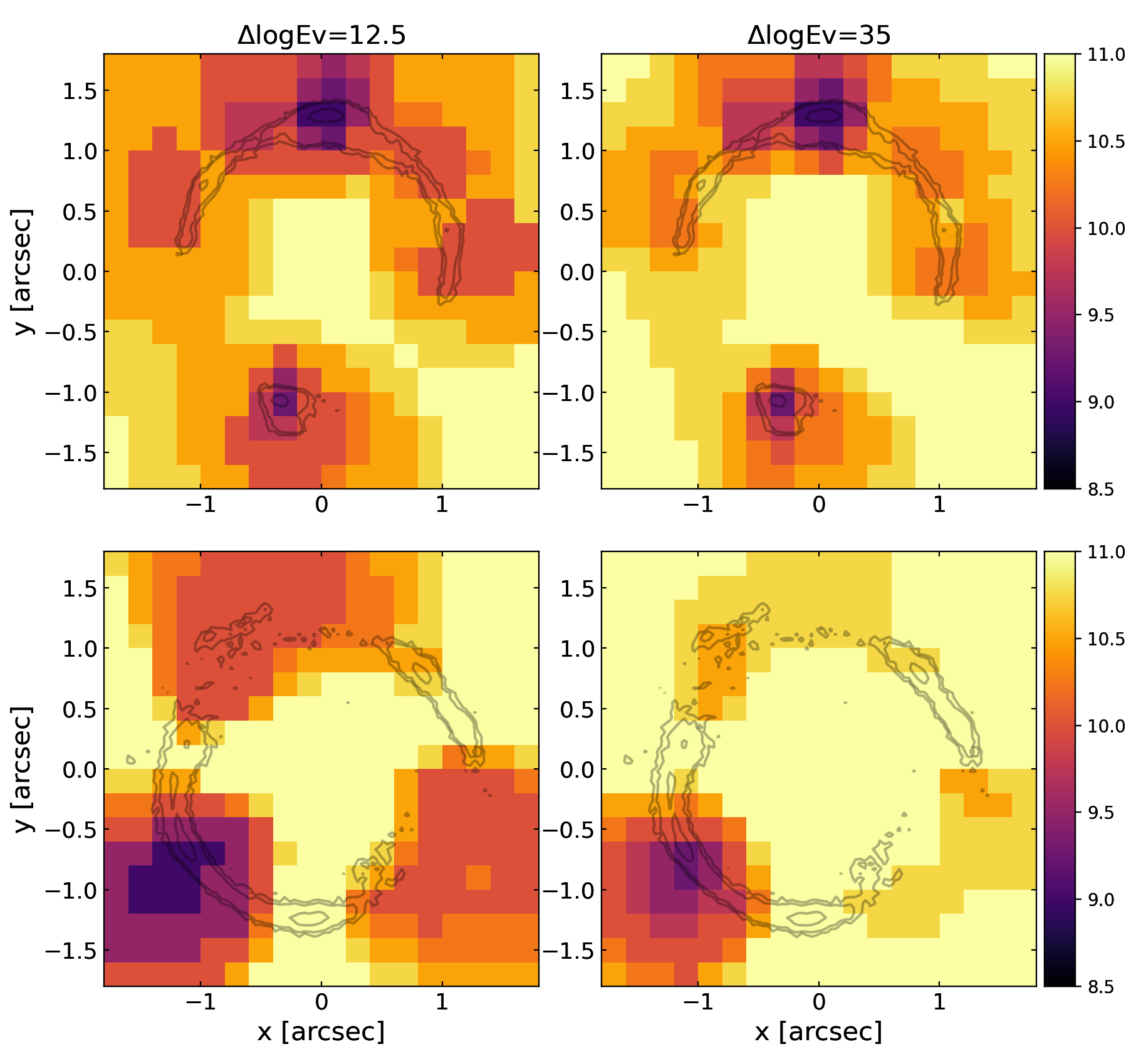}
    \caption{Sensitivity maps of lenses BELLS1 and BELLS2, calculated using a lower evidence threshold with respect to the fiducial one: $\Delta\log$Ev=12.5 on the left and $\Delta\log$Ev=35 on the right. These maps can then be compared with the corresponding ones in Figure \ref{images_hst}.
    }
    \label{sens_lowdelta}
\end{figure}

We used the procedure described in Section \ref{method} to calculate two additional sensitivity maps for the lenses BELLS1 and BELLS2. Instead of the fiducial value $\Delta\log E$=50, adopted throughout the paper, we use two lower evidence threshold: 12.5 and 35. The resulting maps are shown in Figure \ref{sens_lowdelta} and these can be compared with the corresponding ones from Figure \ref{images_hst}: as expected, lowering the threshold produces larger sensitive regions. It is relevant to discuss how the number of expected detections depend on the chosen threshold, since other works \citep{hezaveh16} have claimed subhalo detections with a threshold lower than 50. For these maps the mean value of $M_{\rm low}$ for $\Delta\log$Ev= (12.5,35,50) is (10.32,10.43,10.6) for BELLS1 and (10.1,10.42,10.8) for BELLS2. The minimum $M_{\rm low}$ value is instead (9,9,9.2) and (9.,9.25,9.375).

\bsp	
\label{lastpage}
\end{document}